\documentclass[10pt]{iopart}
\usepackage{epsf}
\usepackage{epsfig}
\def\e{\mbox{e}}
\begin{document}

\title[$\gamma^{(*)} \gamma^{(*)}$ reactions at high energies]{\boldmath
$\gamma^{(*)} \gamma^{(*)}$ \unboldmath reactions at high 
energies\footnote{submitted to proceedings of the
Durham Collider Workshop, Durham, UK, 22-26 September 1999}}

\author{A Donnachie\dag\ and S S\"oldner-Rembold\ddag}

\address{\dag\ Department of Physics, Manchester University, UK, ad@a13.ph.man.ac.uk}

\address{\ddag\ CERN, Geneva, Switzerland, stefan.soldner-rembold@cern.ch}

\begin{abstract}
The energy available for $\gamma^{(*)}\gamma^{(*)}$ physics at LEP2 is
opening a new window on the study of diffractive phenomena, both 
non-perturbative and perturbative. We discuss some of the uncertainties
and problems connected with the experimental measurements and their
interpretation. 
\end{abstract}



\vspace{-3mm}
\section{Introduction}
Diffractive phenomena occur in each of 
untagged, single-tagged and double-tagged photon-photon reactions via the total
hadronic $\gamma\gamma$ cross-section, $\sigma_{\gamma\gamma}$; the 
structure function of the real photon, $F_2^{\gamma}$ (or equivalently 
the $\gamma^{*}\gamma$ cross-section); and the total hadronic $\gamma^{*}
\gamma^{*}$ cross-section, $\sigma_{\gamma^*\gamma^*}$ respectively.
Thus in principle it is possible to study diffraction continuously
from the quasi-hadronic regime dominated by non-perturbative physics
to the realm of perturbative QCD with either single or double hard
scales. 
\section{\boldmath $\gamma\gamma$ \unboldmath scattering}
\begin{figure}[htb]
\vspace{-3mm}
\begin{center}
\mbox{
\epsfxsize=0.85\textwidth
\epsffile{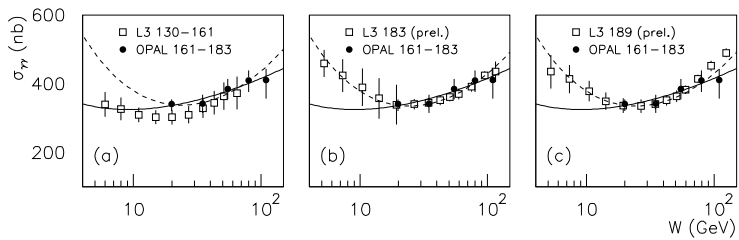}}
\caption{OPAL data at $\sqrt{s}_{\rm ee}=161-183$~GeV
in comparison to the L3 data 
at $\sqrt{s}_{\rm ee}=130-161$~GeV (a), at $\sqrt{s}_{\rm ee}=183$~GeV (b) and 
at $\sqrt{s}_{\rm ee}=189$~GeV (c). In all cases PHOJET was
used for the unfolding of detector effects. The continuous line
is a fit to the OPAL data, the dashed line a fit to
the L3 data at $\sqrt{s}_{\rm ee}=183-189$~GeV (see text).
\label{fig1}}
\end{center}
\end{figure}

The total hadronic $\gamma\gamma$ cross-section was measured at LEP
in the ranges $5\le W \le 145$~GeV~\cite{L397,L399} and 
$10\le W\le 110$~GeV~\cite{OPAL99}, 
where $W$ is the photon-photon centre-of-mass energy (Fig.~\ref{fig1}).
Since the use of different Monte Carlo models 
(PYTHIA~\cite{bib-pythia} or PHOJET~\cite{bib-phojet})
for the unfolding of detector effects leads to significant
shifts of the normalisation, the results of the two
experiments are compared using only PHOJET for these 
corrections\footnote{The published OPAL data are given
after averaging the PHOJET and the PYTHIA corrected 
results (Fig.~\protect\ref{sd1}).}.

Both experiments have measured the high energy rise of the total
cross-section typical for hadronic interactions. However a faster
rise of the total $\gamma \gamma$ cross-section with $W$ compared
to purely hadronic interactions has not been unambiguously observed.
This faster rise is predicted by most models for $\gamma\gamma$ interactions.

To quantify this effect, both experiments have fitted
a Donnachie-Landshoff parametrisation of the form
$\sigma_{\gamma\gamma}=X s^{\epsilon}+ Ys^{-\eta}$ with $\eta=0.34$.
The results are $\epsilon=0.10\pm 0.02$~(OPAL) and
$\epsilon=0.22\pm 0.02$~(L3), where the L3 fit uses
only the preliminary data at $\sqrt{s}_{\rm ee}=183-189$~GeV.
The fitted curves are also shown in Fig.~\ref{fig1}. 
The L3 result implies a significantly faster rise
of the total $\gamma\gamma$ cross-section than in
hadron-hadron scattering whereas the OPAL result
is consistent with a typical value of $\epsilon\approx 0.08$
for a soft Pomeron.

The results are consistent
in the kinematic region where the measurements of both experiments
overlap. Some discrepancies seem to exist between the 
L3 measurements at different $\sqrt{s}_{\rm ee}$, both at low
and at high $W$. The data in the range $W<10$~GeV also have large influence
on the fitted value of $\epsilon$ due to the large correlation
between the Reggeon-Term $Y$ and $\epsilon$.

\begin{figure}[htb]
\begin{tabular}{ccc}
\epsfig{file=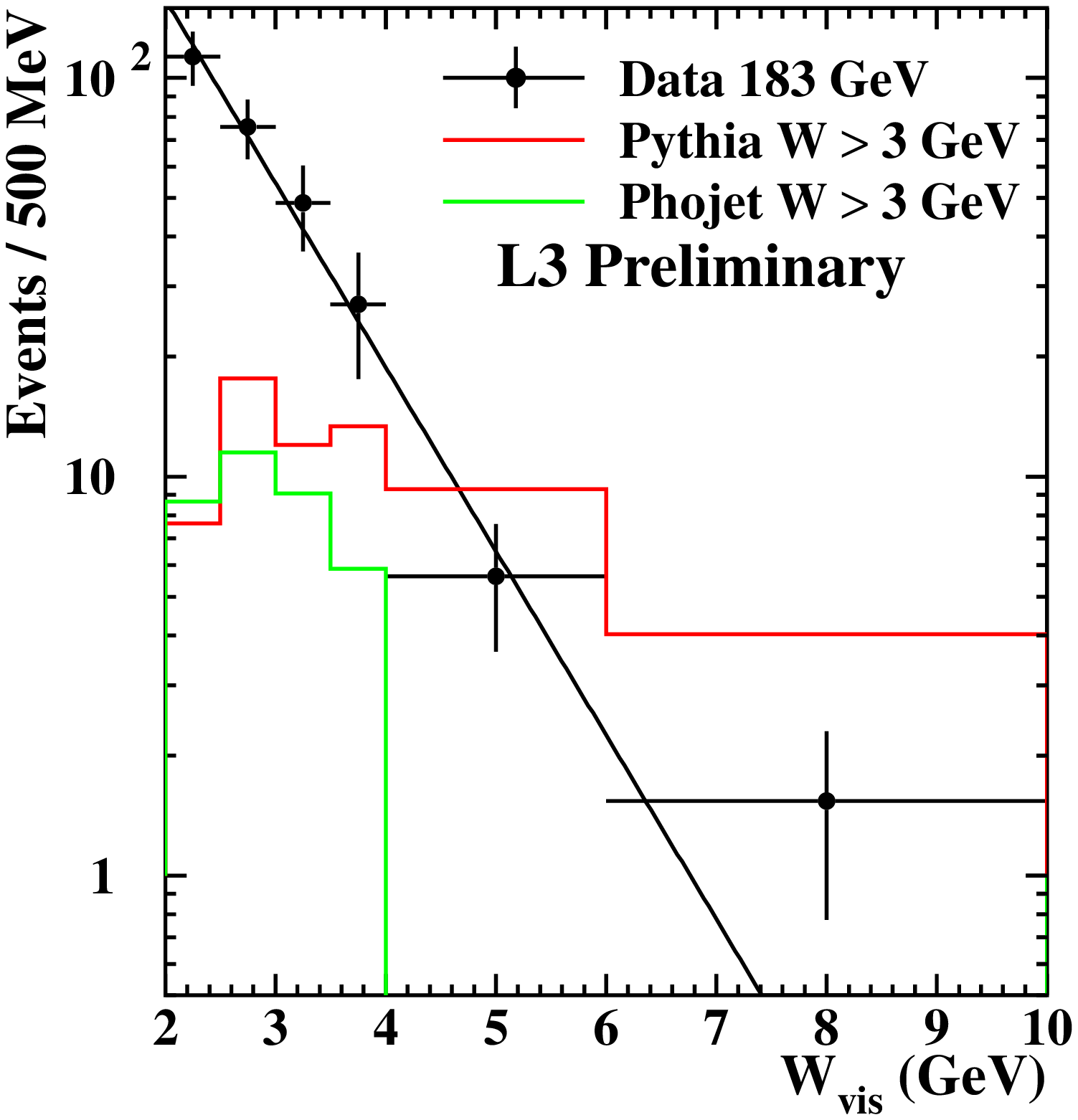,width=0.3\textwidth,height=4cm} &
\epsfig{file=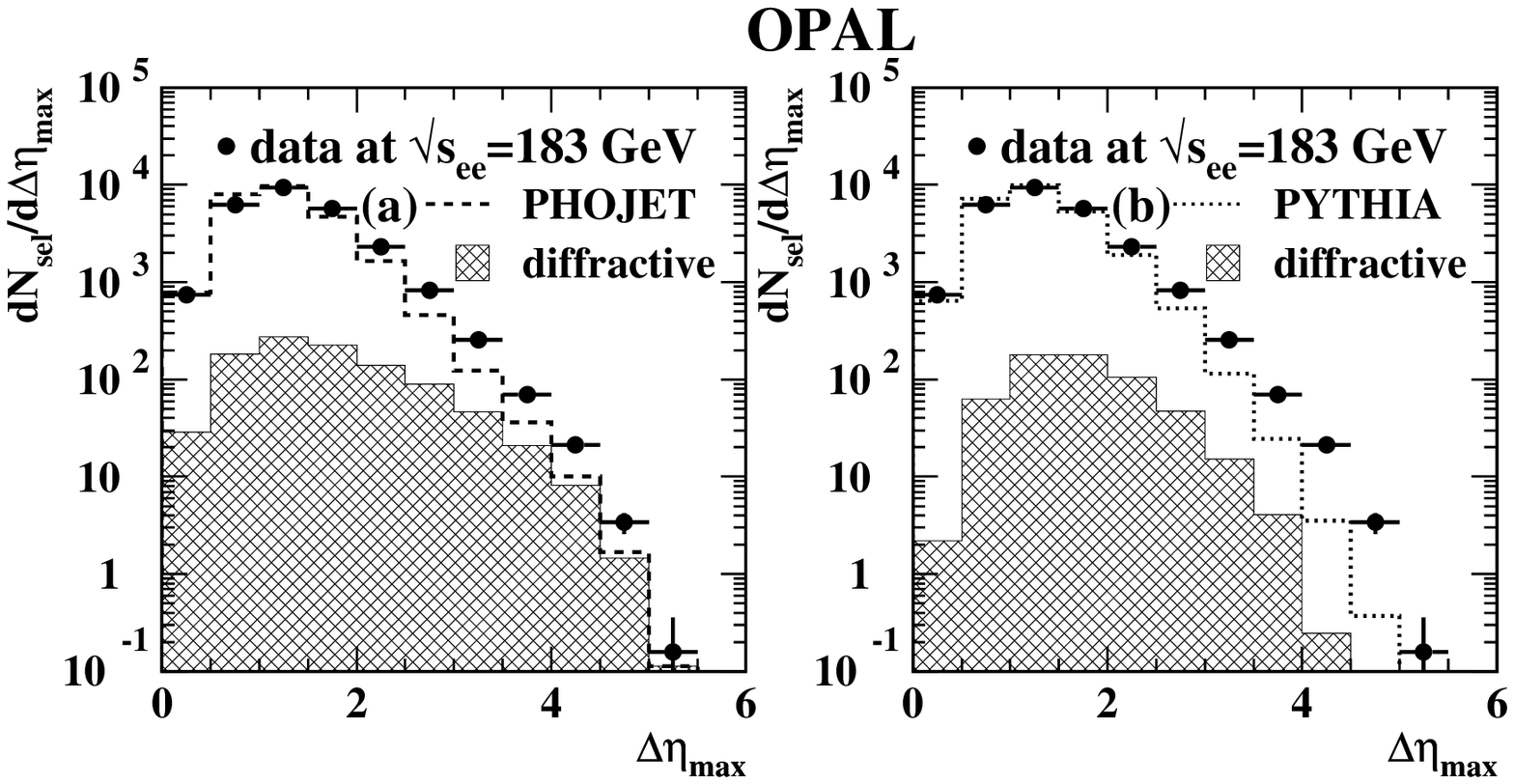,width=0.6\textwidth,height=4cm} &
\end{tabular}
\caption{
Left plot (L3): 
number of $\gamma\gamma\to\rho^0 X$ events as a function
of the visible $W$ ($W_{\rm vis}$). 
Right plots (OPAL):
number of selected two-photon events in the range $6<W_{\rm vis}<120$~GeV
as a function of the
maximum rapidity gap, $\Delta\eta_{\rm max}$, between any two 
particles (tracks and calorimeter clusters). For the MC, the
diffractive events (including quasi-elastic scattering) are shown
separately.} 
\label{fig2}
\end{figure}

The main problems of this measurement are the resolution effects
in the reconstruction of $W$ from the hadronic final state and
the small acceptance for events coming from soft diffractive or
quasi-elastic processes (e.g. $\gamma\gamma\to\rho\rho$) which
lead to the large model dependence for the final results. 
The $W$ resolution makes it necessary
to use unfolding models which introduce large bin-to-bin correlations.
The acceptance for soft diffractive and quasi-elastic processes
is only 5-15$\%$, depending on the $W$ range and on the 
MC model used~\cite{OPAL99}. For $W>20$~GeV the average polar
angle of the pions in $\gamma\gamma\to\rho\rho$ events is less than 100 mrad,
well below the tracking coverage of the LEP detectors. 
L3 has therefore measured inclusive $\rho$ production in $\gamma\gamma$ events
for $3\le W\le 10$~GeV. In this $W$ region large discrepancies
between the Monte Carlo models and the data are observed (Fig.~\ref{fig2}).
At higher $W$ OPAL has studied the maximum rapidity gap 
$\Delta\eta_{\rm max}$ between any two particles (tracks and calorimeter 
clusters) in an event. At high $\Delta\eta_{\rm max}$, where diffractive
events are expected to contribute, the data lie above the Monte Carlo models
and PHOJET is closer to the data than PYTHIA.

Soft processes like quasi-elastic scattering
($\gamma\gamma\to VV$, where $V$ is a vector meson), 
single-diffractive scattering ($\gamma\gamma\to VX$, where $X$ is
a low mass hadronic system) or double-diffractive scattering 
($\gamma\gamma\to X_1 X_2$) are modelled by both generators.
The cross-sections are obtained by fitting
a Regge parametrisation to pp, $\mbox{p}\overline{\mbox{p}}$ 
and $\gamma$p data
and by assuming Regge factorisation, i.e.~universal
couplings of the Pomeron to the
hadronic fluctuations of the photon. 
In both generators
the quasi-elastic cross-section is about $5-6\%$,
the single-diffractive cross-section about $8-12\%$
and the double-diffractive cross-section about $3-4\%$ of 
$\sigma_{\gamma\gamma}$ for $W>10$~GeV.
In the $\gamma\gamma$ data no clear diffractive signal has yet 
been observed and
it would be very useful to find experimental variables which could
give a better discrimination between diffractive and non-diffractive events 
at LEP and which could be used to test the Monte Carlo models.

\section{The hard Pomeron model and the dipole formalism}

As the energies and virtualities available at LEP are comparatively moderate
it is necessary to take into account diffractive and 
non-diffractive contributions. There are 
two main sources of the non-diffractive contributions: 
Reggeon exchange and the quark box diagram with pointlike 
couplings of the photon. In Regge language the latter gives rise to a fixed
pole in the complex angular momentum plane, so it is not dual to Regge 
exchange and it is correct to add the two contributions. The box diagram is 
well defined~\cite{box}. The Regge contribution to $F_2^{\gamma}$ can be 
estimated~\cite{DDR98} using the DGLAP evolved pion structure function and 
naive VMD. This can be extended to both the $\gamma\gamma$ and $\gamma^*
\gamma^*$ cross-sections assuming factorization \cite{DDR99}. In the dipole
approach to the small-$x$ structure function of the proton it has become 
increasingly clear that the nominally perturbative regime still contains 
some non-perturbative contribution \cite{GW99,FKS99,FGMS99}. A specific
model has been proposed in terms of two Pomerons \cite{DL98,DL99,PVL99}.
This combines a hard Pomeron with an intercept of about 1.44 together 
with the soft Pomeron of hadronic physics with an intercept of about 1.08.

An analogous approach is that of~\cite{ADMNR} 
in which the hard Pomeron is modelled within the BFKL
framework. A similar conclusion is reached, that for diffractive 
reactions on a hadronic target the purely perturbative regime is not
reached until rather large values of $Q^2$.

Combining the dipole formalism with the two-Pomeron approach allows 
predictions to be made for the $\gamma^{(*)}\gamma^{(*)}$ cross-sections
\cite{DDR99}. An appropriate model for soft Pomeron exchange is the eikonal 
approach \cite{Nac91} to high energy scattering. It is particularly suited to 
incorporate the non-perturbative aspects of QCD which are treated in the 
Model of the Stochastic Vacuum~\cite{Dos87,DS88,DFK94}, which approximates 
the infrared part of QCD by a Gaussian stochastic process in the colour 
field strength. The two-Pomeron approach of~\cite{DL98} has been adapted 
to the MSV model in \cite{Rue98} and successfully tested for the photo- 
and electroproduction of vector mesons, and for the proton structure 
function over a wide range of $x$ and $Q^2$.
\begin{figure}[htb]
\begin{tabular}{ccc}
\epsfig{file=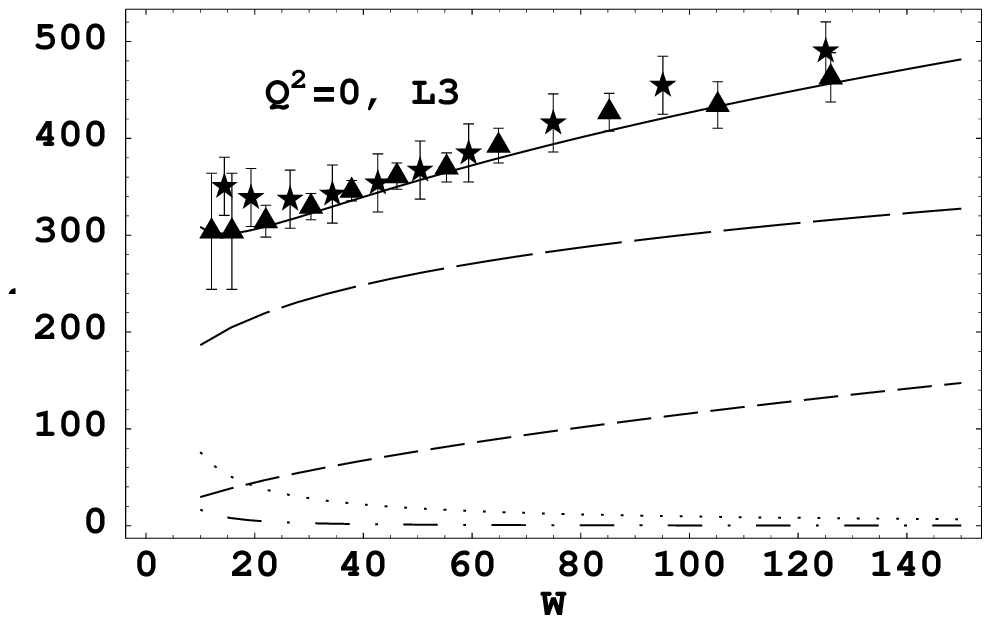,width=0.45\textwidth} &
\epsfig{file=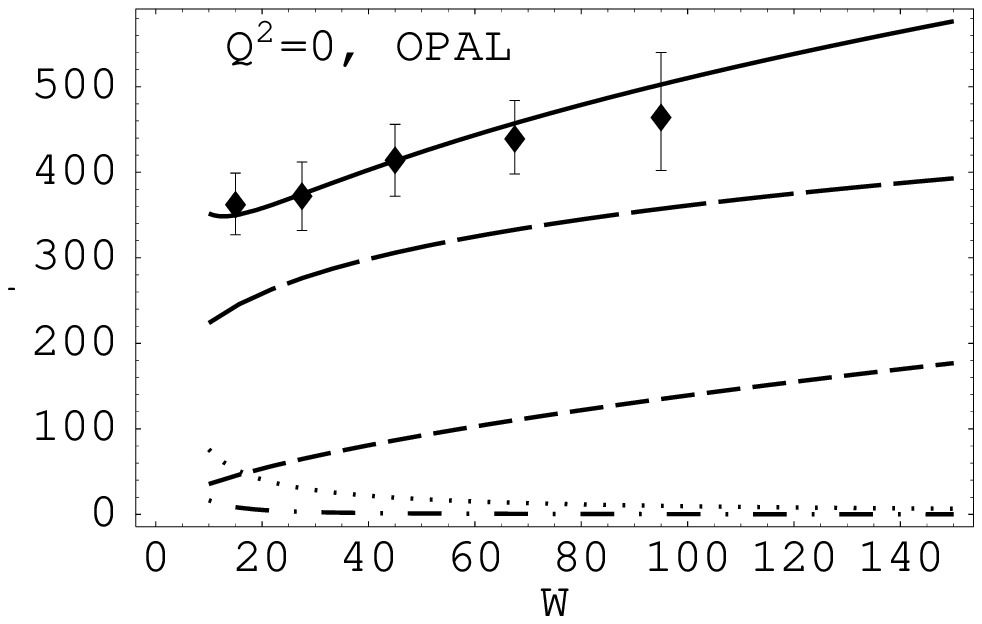,width=0.45\textwidth} &
\end{tabular}
\caption{The total hadronic photon-photon cross-section 
$\sigma_{\gamma\gamma}$ (in nb) as function of $W$ (in GeV).}
\label{sd1}
\end{figure}
\begin{figure}[htb]
\begin{tabular}{ccc}
\epsfig{file=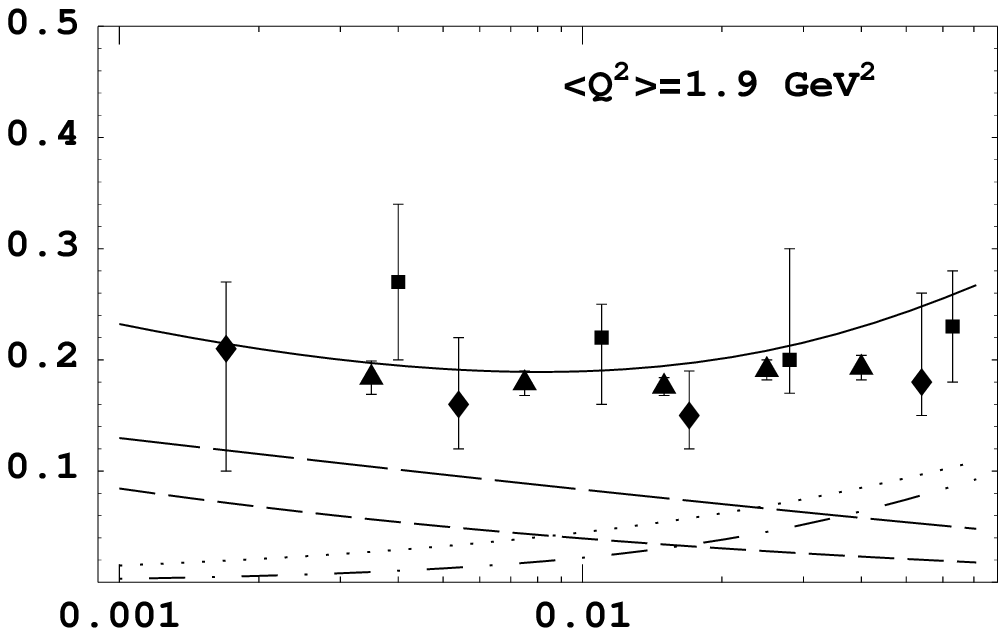,width=0.45\textwidth} &
\epsfig{file=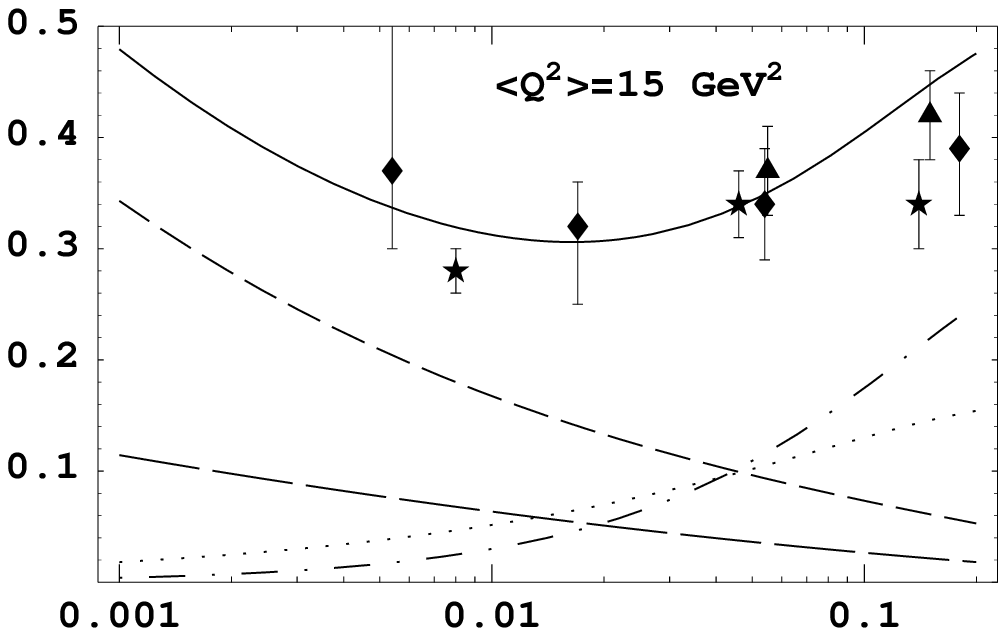,width=0.45\textwidth} &
\end{tabular}
\caption{The photon structure function $F_2^\gamma$ as 
function of $x$. 
a) $\langle Q^2\rangle =1.9$~GeV$^2$,
(L3: triangles, OPAL: diamonds and boxes);
b) $\langle Q^2\rangle \approx 15$~GeV$^2$,
(ALEPH: stars, L3: triangles, OPAL: diamonds and boxes).}
\label{sd2}
\end{figure}
\begin{figure}[htb]
\begin{tabular}{ccc}
\epsfig{file=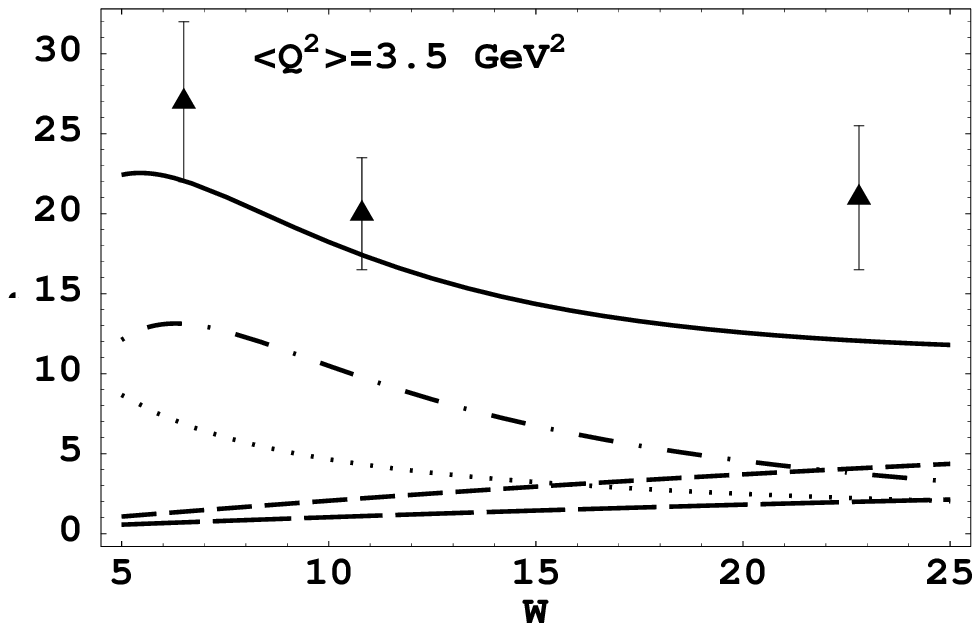,width=0.45\textwidth} &
\epsfig{file=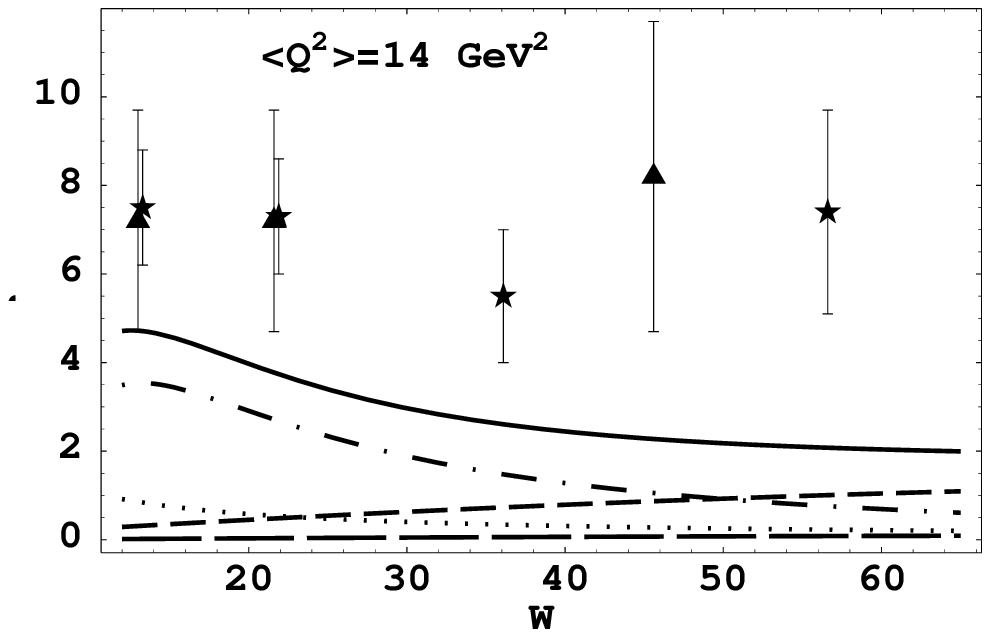,width=0.45\textwidth} &
\end{tabular}
\caption{The virtual photon cross-section $\sigma_{\gamma^*\gamma^*}$ in nb.
L3 data points are shown.}
\label{sd3}
\end{figure}

With all parameters determined from hadronic scattering and deep inelastic 
scattering, in principle the model can then predict all the $\gamma^{(*)}
\gamma^{(*)}$ cross-sections~\cite{DDR99}. The only caveat is that the Pomeron
contribution to $\sigma_{\gamma\gamma}$ is rather sensitive to the effective 
light quark mass $m_{\rm q}$ entering the photon wave function, varying as 
$\sim 1/m_{\rm q}^4$. This is illustrated in Figs.~\ref{sd1}a and b which show 
separately the L3~\cite{L397,L399} and OPAL~\cite{OPAL99} data
and the Pomeron model with $m_{\rm q} = 210$ MeV and 200 MeV respectively,
together with the other contributions to the total cross-section. These values
of $m_{\rm q}$ are within the range previously determined, and the choice does not 
affect the predictions away from $Q^2 = 0$. Comparison of the predictions with
$F_2^\gamma$ at $\langle Q^2 \rangle = 1.9$ and $15$ GeV$^2$ are shown in 
Figs.~\ref{sd2} a and b respectively. They clearly provide a satisfactory 
description of the data. However comparison with $\sigma_{\gamma^*\gamma^*}$
is much less successful as is evident in Figs.~\ref{sd3}a and b, for which
$\langle Q_1^2 \rangle = \langle Q_2^2 \rangle = 3.5$ and 14 GeV$^2$
respectively.

The significance of these results is that a well-tried model of diffraction
which successfully describes high-energy hadronic interactions, vector meson
photo- and electroproduction, deep inelastic scattering at small $x$, the 
real $\gamma\gamma$ cross-section and the structure function of the real
photon fails to predict correctly the $\gamma^*\gamma^*$ cross-section even
at quite modest photon virtualities. This is clearly due to the fact that, 
uniquely among these various processes, the $\gamma^*\gamma^*$ interaction
involves two small dipoles. It emphasizes the importance of the $\gamma^*
\gamma^*$ cross-section as a probe of the dynamics of the perturbative hard 
Pomeron. 

\section{\boldmath $\gamma^{*}\gamma^{*}$ \unboldmath scattering in
the BFKL formalism}
The application of the BFKL formalism to $\gamma^*\gamma^*$ scattering
has been considered by [21-26].
In the BFKL formalism there is a problem at LLO in setting the two mass
scales on which the cross-section depends: the mass $\mu^2$ at which the 
strong coupling $\alpha_s$ is evaluated and the mass $Q_s^2$ which provides 
\begin{figure}[htb]
\begin{tabular}{ccc}
\epsfig{file=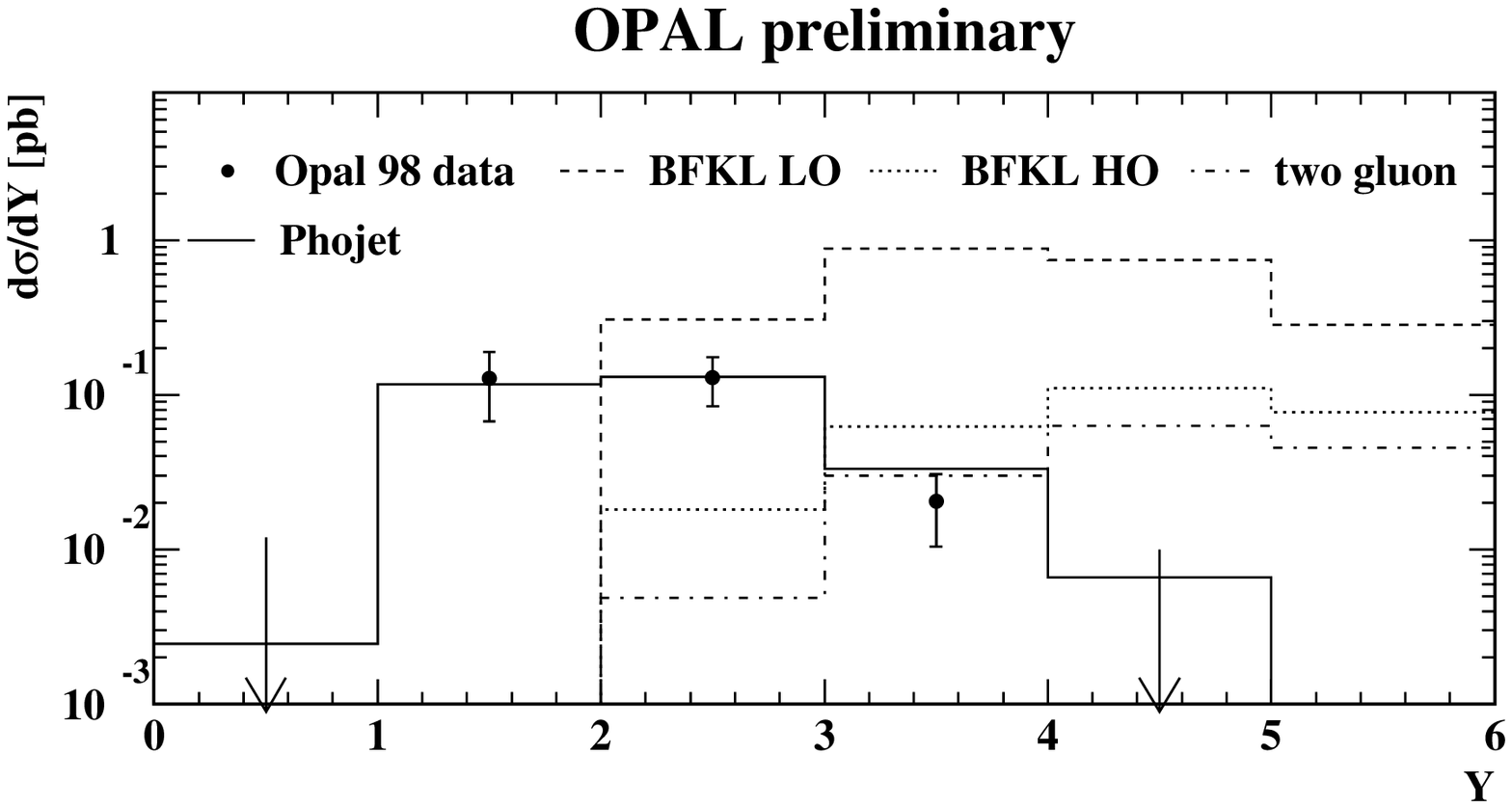,width=0.476\textwidth} &
\epsfig{file=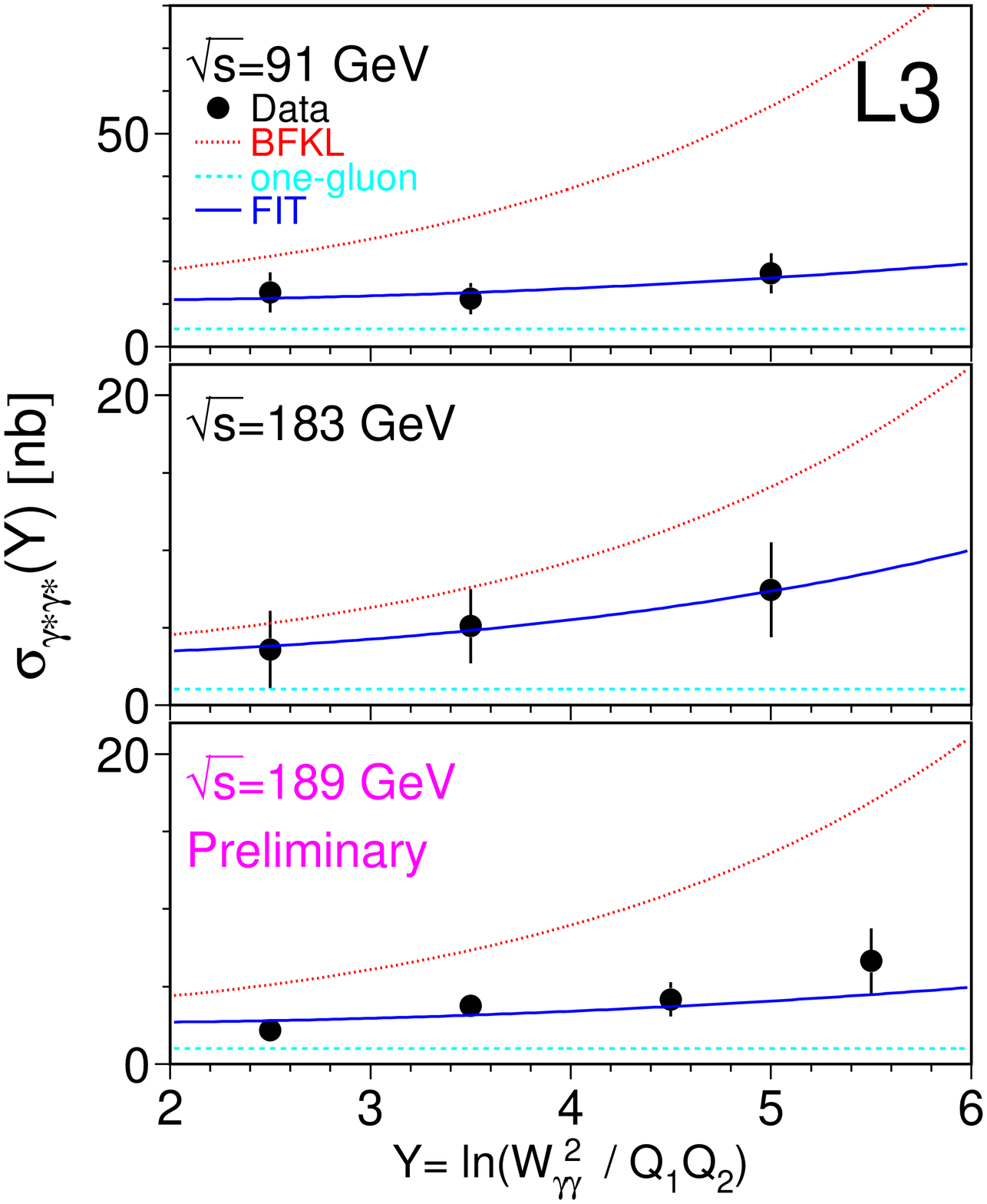,width=0.476\textwidth,height=5cm} &
\end{tabular}
\caption{Cross-section for the process
$\e^+\e^-\to \e^+\e^-\gamma^*\gamma^* \to \e^+\e^-$~hadrons
as a function of $Y\approx \ln (W^2/Q_1^2 Q_2^2)$ (for exact
kinematic cuts see~\protect\cite{OPAL00} and \protect\cite{L3dt}).
}
\label{dt}
\end{figure}
the scale for the high energy logarithms. The result is very sensitive to 
these parameters, and Brodsky et al \cite{bhs0,bhs} showed that changing 
$\mu^2 \to 4\mu^2$ or $Q_s^2 \to Q_s^2/4$ alters the predicted cross-section 
by factors of $\sim 1/4$ or $\sim 4$ respectively in a typical LEP experiment.
An additional uncertainty is due to the correct treatment of the production
of massive charm quarks.

In an attempt to overcome the scale problem, Boonekamp et al \cite{bdrw} take 
a phenomenological approach to estimate the NNLO effects, making use of a fit
\cite{bfklf} to the proton structure function using the QCD dipole picture 
of BFKL dynamics. This reduces both the size of the BFKL cross-section and 
its energy dependence. Fig.~\ref{dt} shows the preliminary OPAL 
measurement of the double-tag e$^+$e$^-$ cross-section in
the range $Q_1^2\approx Q_2^2 \approx 5-25$~GeV$^2$~\cite{OPAL00} 
compared to the LO
BFKL calculation and to the HO model of~Boonekamp et al.~\cite{bdrw}. 
The cross-section predicted by PHOJET~\cite{bib-phojet} is also shown. 
The L3 collaboration has
extracted the $\gamma^*\gamma^*$ cross-section using the photon flux
for transverse photons (Fig.~\ref{dt})~\cite{L3dt}. The QPM part (box diagram)
has been subtracted (the unsubtracted cross-section is shown in
Fig.~\ref{sd3}). 
The L3 data is compared to a LO BFKL prediction and to
the two-gluon exchange cross-section (here called one-gluon) based on
Ref.~\cite{bhs} and to a fit of the hard Pomeron intercept.
A calculation~\cite{jan} of subleading
corrections to the BFKL equation shows that these are significant at LEP
energies, and with the inclusion of the soft Pomeron a reasonable
description of the L3 data is obtained.

Both experiments observe that the cross-section predicted by
PHOJET, which does not
contain BFKL effects, is consistent with the data within the large
experimental errors,
whereas LO BFKL predictions overestimate the $\gamma^*\gamma^*$ cross-section
by a large factor. However, the large theoretical 
uncertainties discussed above need
to be taken into account.

\section{Conclusions}
In the last year difficulties have emerged with the application of the
Altarelli-Parisi equation to the evolution of the proton structure
function at small $x$ and with the BFKL equation. These are summarised 
in~\cite{BL}. One of the questions is whether intrinsically non-perturbative
contributions are involved, even at quite large $Q^2$, because of the
intrinsically non-perturbative target. This complication is in principle 
avoided in $\gamma^*\gamma^*$ reactions as both are dominated at large
$Q^2$ by the perturbative part of the photon wave function. 
This may be happening at quite modest values of
$Q^2$, providing LEP with an excellent opportunity to clarify this
question.

\end{document}